\documentclass[debug]{rmaa}

\usepackage{paralist}

\usepackage{psfrag,color}

\usepackage[latin1]{inputenc}

\title{Chromospherically Active \\
Low-Mass Close Binary: KIC 9761199}

\author{E. YOLDA\c{S}\altaffilmark{1} and H. A. DAL\altaffilmark{1,2}}

\altaffiltext{1}{Department of Astronomy and Space Sciences, University of Ege, Bornova, 35100 ~\.{I}zmir, Turkey}
\altaffiltext{2}{Corresponding Author, Email: ali.dal@ege.edu.tr}

\shortauthor{Yolda\c{s} E.\& Dal, H. A.}
\shorttitle{RevMexAA Main Journal Demo Document}

\fulladdresses{
\item Department of Astronomy and Space Sciences, University of Ege, Bornova, 35100 ~\.{I}zmir, Turkey (HAD, ali.dal@ege.edu.tr, EY, ezgiyoldas@gmail.com).}

\listofauthors{Yolda\c{s} E.\& Dal, H. A.}
\indexauthor{Yolda\c{s}, E.}
\indexauthor{Dal, H. A.}

\abstract{In this study, we present the results obtained from KIC 9761199's the photometrical data acquired by the Kepler Mission. The light curve of the system, the sinusoidal variation out-of-eclipses and instant-short term flare events in the entire light curves were analyzed. The temperature of the secondary component was found to be 3891$\pm$1 K, while the mass ratio of the components ($q$) was found to be 0.69$\pm$0.01, and the orbital inclination ($i$) was computed as $77^\circ.4\pm0^\circ.1$. The sinusoidal variation is caused by the stellar spots of two active regions separated by about $180^\circ$ longitudinally located around the latitudes of $+47^\circ$ and $+30^\circ$. In addition, 94 flares were detected and their parameters were computed. The OPEA model was derived for these flares and its parameters were computed. The $Plateau$ value as saturation level for the active component was found to be 1.951$\pm$0.069 s, while the $half-life$ value was found to be 1014 s. The flare frequency $N_{1}$ was found to be 0.01351 $h^{-1}$, while the flare frequency $N_{2}$ was found to be 0.00006. Maximum flare rise time ($T_{r}$) was found to be 1118.098 s, while maximum flare total time ($T_{t}$) was found to be 6767.72 s. Comparing its analogue it is seen that the chromospheric activity level of KIC 9761199, which is a low-mass close binary system according to the light curve analyses, is an expected level according to the ($B-V$) color index of $1^{m}.303$ for the active component.}

\addkeyword{techniques: photometric}
\addkeyword{methods: data analysis}
\addkeyword{stars: binaries: eclipsing}
\addkeyword{stars: low-mass}
\addkeyword{stars: flare}
\addkeyword{stars: individual: KIC 9761199}

\begin{document}
\maketitle

\section{Introduction}
\label{sec:intro}
It is well known for several decades that the population rate of red dwarfs in our Galaxy is about 65$\%$, while seventy-five percent of them show flare activity. The stars exhibiting flare activity are known as UV Ceti type stars \citep{Rod86}. As it is seen that the population rate of UV Ceti type stars in our Galaxy is about 48.75$\%$, which means that one of each two stars in our galaxy shows the flare phenomenon. As it is expected, the general population rate of UV Ceti type stars is incredibly high in both the open clusters and the associations \citep{Mir90, Pig90}. However, the population rate of these stars has a reducing trend, while the age of the cluster gets older due to the Skumanich's law \citep{Sku72, Pet91, Sta91, Mar92}. This is because the mass loss rate is so high due to the flare activity during the main sequence life of the stars that the evolution track of the stars are changed by these mass loss in their main sequence stages, and also in later stages \citep{Mar92}. 

According to the recent studies, the mass loss rate of UV Ceti type stars is about $10^{-10}$ $M_{\odot}$ per year due to flare like events, while the solar mass loss rate is about $2\times10^{-14}$ $M_{\odot}$ per year \citep{Ger05}. The difference between the mass loss ratios is also seen between the flare energy levels of the solar and stellar cases. As it is generally observed in the case of RS CVn type active binaries \citep{Hai91}, the highest energy detected from the most powerful flares, known as two-ribbon flares, occurring on the sun is found to be $10^{30}$ - $10^{31}$ erg \citep{Ger05, Ben08}. On the other hand, the flare energy level varies from $10^{28}$ erg to $10^{34}$ erg in the case of dMe stars \citep{Hai91, Ger05}, while some stars of young clusters such as the Pleiades cluster and Orion association exhibit some powerful flare events, which energies reach $10^{36}$ erg \citep{Ger83}.

Although both the mass loss ratios and the flare energy levels are remarkably different between the solar and stellar cases, the flare events occurring on a dMe star are generally tried to explain by the classical theory of solar flare. The main energy source in the flare events is magnetic reconnection processes in this theory \citep{Ger05, Hud97}. However, this theory does not figure out each flare phenomenon observed from the UV Ceti type stars. To reach the real solution, examining the flare events occurring on the different type stars, all the differences and similarities should be demonstrated. Then, it should be identified which parameters such as singularity, binarity, mass, age, ect, cause these differences and similarities. For example, some differences are seen between the flare frequencies from one star to the next. The similar case occurs for the flare energy spectra.

In this study, we figured out the nature of a low mass eclipsing binary, KIC 9761199, which is a bit different from the classical UV Ceti type stars from spectral type dMe due to being a binary system. We did complete light curve analyses of the system for the first time in the literature in order to find out the physical properties of the components, using the PHOEBE V.0.32 software \citep{Prs05}, whose method depends on the 2003 version of Wilson-Devinney Code \citep{Wil71, Wil90}. Then, the variations out-of-eclipses were analyzed, and the flares occurring on the chromospherically active component were detected to model the magnetic activity nature of the system, comparing the active component with its analogue.

KIC 9761199 (KOI 1459) was listed with the B band brightness of $17^{m}.2$ in the USNO ACT Catalog \citep{Urb97} in the first time literature. Secondly, KIC 9761199 was listed as J19083435+4630290 in 2MASS All-Sky Survey Catalogue, in which the JHK brightness of the system were given as $J=13^{m}.574$, $H=12^{m}.926$, $K=12^{m}.782$ \citep{Kha01, Cut03}. Although the orbital period of the system ($P_{orb}$) was determined as $0.692031$ day for the first time by \citet{Wat06}, analyzing the data obtained in the Kepler Mission, \citet{Cou11} indicated that the orbital period of the system ($P_{orb}$) is $1.3839980$ day. There is no any light curve analysis of the system in detail, though \citet{Bor11} computed the semi-major axis ($a$) of the system as 0.013 $AB$, while \citet{Cou11} calculated the inclination ($i$) of the system as $74^\circ.47$ and noted that there is no third light excess. However, they also noted that the system should be examined by a complete light curve analysis in detail.

In the Kepler Database, the ($B-V$) color index of the system is given as $0^{m}.068$, while the temperature ratio of the components is given 0.682 by \citet{Sla11}. In addition, the inclination ($i$) of the system is listed as $sini=0.99451$. However, this inclination value is not in agreement with $i=74^\circ.47$ given by \citet{Cou11}. According to \citet{Wal13}, the age of the system is 0.77 Gyr and the ($B-V$) color index of the system is $1^{m}.36$.

The one of the spectral studies of the system is belong to \citet{Man13} and they gave the luminosity ($L$) of the system as 0.082 $L_{\odot}$. Using the calibration of some spectral observations, \citet{Mui12} indicated that the system is a binary with components from the spectral type of M1, whose distance is about 198 pc. The total masses given in the literature for the components vary from 0.51 $M_{\odot}$ \citep{Mui14} to 0.65 $M_{\odot}$ \citep{Cou12}, while the radius vary from 0.48 $R_{\odot}$ \citep{Mui14} to 0.84 $R_{\odot}$ \citep{Cou12}. The temperature given in the literature for the system varies from 3742 K \citep{Mui14} to 4060 K \citep{Cou12}. Using the calibrations, \citet{Cou12} calculated mass and radius of each component as $M_{1}=0.646$ $M_{\odot}$, $M_{2}=0.444$ $M_{\odot}$, $R_{1}=0.830$ $R_{\odot}$, $R_{2}=0.384$ $R_{\odot}$.

\section{Data and Analyses}
\label{sec:Data and Analyses}
The Kepler Mission is one of the space missions to be aimed to find out exo-planet, in which More than 150.000 targets have been already observed \citep{Bor10, Koc10, Cal10}. The quality and sensitivity of these observations have the highest one ever reached in the photometry \citep{Jen10a, Jen10b}. In these observations, so many variable targets such as new eclipsing binaries, etc. have been also discovered apart from the exo-planets \citep{Sla11, Mat12}. There are lots of single or double stars, which some of them are the eclipsing binaries, exhibiting chromospheric activity among these newly discoveries \citep{Bal15}.

The data analyzed in this study were taken from the Kepler Mission Database \citep{Sla11, Mat12}. After removing the all the observations with large error due to the technical problems from the data, using the ephemeris taken from the Kepler Mission database, the phases were computed for all data used in the analysis, and obtained light curves were shown in Figure 1. Because of this study format, the detrended short cadence data were used in the analysis instead of the long cadence. The data were arranged as the suitable formats to analysis the flare events and also the light curve.

\subsection{Light Curve Analysis}
To compute the flare parameters, it needs the synthetic light curve as assumed the quiescent levels. In this purpose, the light curves shown in the upper panel of Figure 1 were analyzed, after removing the instant-short term light variations due to the flare activity. The initial analyses indicated that there are several solutions according to these data. Because of this, the averages of all the detrended short cadence data were computed phase by phase with interval of 0.001. Then, this averaged light curve shown in Figure 2 was analyzed.

Using the PHOEBE V.0.32 software \citep{Prs05}, which is employed in the 2003 version of the Wilson-Devinney Code \citep{Wil71, Wil90}, we analyzed the light curves obtained from the averages of all the detrended short cadence data. We attempted to analyze the light curves in various modes, including the detached system mode (Mod2), semi-detached system with the primary component filling its Roche-Lobe mode (Mod4), and semi-detached system with the secondary component filling its Roche-Lobe mode (Mod5). Our initial analyses demonstrated that an astrophysically reasonable solution was obtainable only in the detached system mode; no results that were statistically consistent with reasonable solutions could be obtained in any of the other modes.

There are several studies about KIC 9761199 in the literature, and lots of temperature values were given for the system varying from 3742 K \citep{Mui14} to 4060 K \citep{Cou12}. To understand which one is right, in this study, we calculated both ($H-K$) and ($J-H$) color indexes from the JHK brightness ($J=13^{m}.574$, $H=12^{m}.926$, $K=12^{m}.782$) given by \citet{Kha01} and \citet{Cut03}. Then, using the calibrations determined by \citet{Tok00}, we derived de-reddened colors as $(H-K)_{\circ}=0^{m}.15$ and $(J-K)_{\circ}=0^{m}.58$. In addition, using the same calibrations, we derived the temperature of the primary component as 4040$\pm$20 K depending on these de-reddened colors. In fact, the initial analyses with different temperature values between 3742 K - 4060 K indicated that an astrophysically acceptable solution could be obtained, if the temperature would be taken 4060 K for the primary component. Because of this, in the analysis, the temperature of the primary component was fixed to 4060 K, while the temperature of the secondary component was taken as adjustable parameter. Considering the spectral type corresponding to this temperature, the albedos ($A_{1}$ and $A_{2}$) and the gravity-darkening coefficients ($g_{1}$ and $g_{2}$) of the components were adopted for the stars with the convective envelopes \citep{Luc67, Ruc69}. The nonlinear limb-darkening coefficients ($x_{1}$ and $x_{2}$) of the components were taken from \citet{Van93}. In the analyses, their dimensionless potentials ($\Omega_{1}$ and $\Omega_{2}$), the fractional luminosity ($L_{1}$) of the primary component, the inclination ($i$) of the system, the mass ratio of the system ($q$), and the semi-major axis ($a$) were taken as the adjustable free parameters.

The sinusoidal variation out-of-eclipses were modelled with two cool spots on the primary component. In the analysis, although the third light was taken as the adjustable free parameters, it is seen that there is no third light excess in the total light.

All the parameters obtained from the light curve analysis are listed in Table 1. As it can be seen from the table, the error values of each parameter are saliently small. This is caused due to the averaged data used in the analyse. If the original data plotted in the upper panel of Figure 1 were used in the analyse, it would be waiting that the error values are more larger. In addition, apart from some theoretical parameters such as the nonlinear limb-darkening coefficients ($x_{1}$ and $x_{2}$), the albedos ($A_{1}$ and $A_{2}$) and the gravity-darkening coefficients ($g_{1}$ and $g_{2}$), the temperature of the primary component is just one parameter to fixed in the analysis. However, the temperature of the primary component was not also arbitrary determined. This temperature was determined from the JHK data of system given in the 2MASS All-Sky Survey Catalogue. As a result, the parameters found from the light curve analysis in Table 1 are absolutely trustworthy to figure out the obvious nature of KIC 9761199.

The synthetic light curve obtained with these parameters is shown in Figure 2. In this figure, both primary and secondary minima of the light curve are also plotted in the wide plane to better view for the readers. In addition, the 3D model of Roche geometry depending on these parameters is shown in Figure 3. The Roche geometry and the cool spot locates depending on the parameters obtained from the light curve analysis is shown for different phases, such as the phases of 0.00, 0.25, 0.50, 0.75.

Although there is not any available radial velocity curve, we tried to estimate the absolute parameters of the components. Considering the calibrations given by \citet{Tok00}, ($B-V$) color index was found to be $1^{m}.303$ for the primary component depending on its temperature value derived from the analysis, while it was found to be $1^{m}.378$ for the secondary components. These color indexes are in agreement with the value found by \citet{Wal13}. Using the calibrations given by \citet{Tok00}, the mass of the primary component must be 0.57$\pm$0.01 $M_{\odot}$ corresponding to its ($B-V$) color index. Considering the possible mass ratio of the system, the mass of the secondary component was found to be 0.39$\pm$0.02 $M_{\odot}$. Using Kepler's third law, we calculated the possible semi-major axis as a 5.16$\pm$0.02 $R_{\odot}$. Considering this estimated semi-major axis, the radius of the primary component was computed as 0.62$\pm$0.05 $R_{\odot}$, while that of the secondary component was computed as 0.56$\pm$0.05 $R_{\odot}$.

\subsection{Flare Activity and the OPEA Model}

To demonstrate the chromospherically active nature of the system, it needs to reveal the light variations just due to flare events. For this aim, first of all, all the light variations due to both the geometrical effects and the rotational modulation caused by the spots were removed from the general light variation. For this purpose, the data of all the pre-whitened light curves were obtained. To acquire the pre-whitened data, we extracted the synthetic light curves from all the detrended short cadence data. 

In the second step, to determine the basic flare parameters such as the first point of the flare beginning, the last point of the flare end and the flare energy, the quiescent levels for each flare should be derived. In this point, the synthetic model lead us to definite the quiescent levels for each flare at the same time of that flare. Using the synthetic model as the quiescent levels, the parameters of the flares were computed. Two samples of the flare light curves taken from observation data and the quiescent levels derived for these flares are shown in Figures 4.

Using the synthetic models assumed as the quiescent levels, both the beginning and the end of a flare for each one were defined, and then, some flare parameters, such as flare rise times ($T_{r}$), decay times ($T_{d}$), amplitudes of flare maxima, flare equivalent durations ($P$), were computed. In total, 94 flare were detected from the available short cadence data in the Kepler Mission Database. All the computed parameters are listed in Table 2 for these 94 flares. In the table, flare maximum times, equivalent durations, rise times, decay times and amplitudes of flare maxima are listed from the first column to the last, respectively.

In the analysis, the equivalent durations of the flares were computed using Equation (1) taken from \citet{Ger72}:

\begin{center}
\begin{equation}
P~=~\int[(I_{flare}~-~I_{0})/I_{0}] dt
\end{equation}
\end{center}
where $I_{0}$ is the flux of the star in the observing band while in the quiet state. In this study, we computed $I_{0}$ using by the synthetic models derived with the light curve analysis. $I_{flare}$ is the intensity observed at the moment of the flare. $P$ is the flare-equivalent duration in the observing band. In this study, the flare energies ($E$) were not computed to be used in the following analyses due to the reasons described in detail by \citet{Dal10, Dal11}.

Examining the relationships of the flare parameters among each other, it is seen that the distributions of flare equivalent durations on the logarithmic scale versus flare total durations are varying according to a rule. The distributions of flare equivalent durations on the logarithmic scale cannot be higher than a specific value for the star, and it is no matter how long the flare total duration is. Using the SPSS V17.0 \citep{Gre99} and GrahpPad Prism V5.02 \citep{Daw04} programs, \citet{Dal10, Dal11} indicated that the best function is the One Phase Exponential Association (hereafter OPEA) for the distributions of flare equivalent durations on the logarithmic scale versus flare total durations. The OPEA function has a $Plateau$ term, and this makes it a special function in the analyses. The OPEA function is defined by Equation (2):

\begin{center}
\begin{equation}
y~=~y_{0}~+~(Plateau~-~y_{0})~\times~(1~-~e^{-k~\times~x})
\end{equation}
\end{center}
where the parameter $y$ is the flare equivalent duration on a logarithmic scale, the parameter $x$ is the flare total duration, according to the definition of \citet{Dal10}, and the parameter $y_{0}$ is the flare-equivalent duration in on a logarithmic scale for the least total duration. In other words, the parameter $y_{0}$ is the least equivalent duration occurring in a flare for a star. Here is an important point that the parameter $y_{0}$ does not depends on only flare mechanism occurring on the star, but also depends on the sensitivity of the optical system used for the observations. The parameter $Plateau$ value is upper limit for the flare equivalent duration on a logarithmic scale. \citet{Dal11} defined $Plateau$ value as a saturation level for a star in the observing band. 

Using the least-squares method, the OPEA model was derived for the distributions of flare equivalent durations on the logarithmic scale versus flare total durations. The derived model is shown in Figure 5 together with the observed flare equivalent durations, while the parameters computed from the model are listed in Table 3. The $span$ value listed in the table is difference between $Plateau$ and $y_{0}$ values. According to the definition of the OPEA function, the parameter $k$ in Equation (2) is a constant for just this model, depending on the x values. The $half-life$ value is half of the first $x$ value, at which the model reaches the $Plateau$ value. In other words, it is half of the minimum flare total time, which is enough to the maximum flare energy occurring in the flare mechanism.

It was tested by using three different methods, such as the D'Agostino-Pearson normality test, the Shapiro-Wilk normality test and also the Kolmogorov-Smirnov test, given by \citet{Dag86} to understand whether there are any other functions to model the distributions of flare equivalent durations on the logarithmic scale versus flare total durations. In these tests, the probability value called as $p-value$ was found to be $p-value < 0.001$ and this means that there is no other function to model the distributions of flare equivalent durations \citep{Mot07, Spa87}.

KIC 9761199 was observed as long as 289.82931 day from JD 2455641.50631 to JD 2455931.33562 without any remarkable interruptions. In total, significant 94 flares were detected in these observations. The total flare equivalent duration computed from all the flares was found to be 628.11671 s (0.17448 hours). \citet{Ish91} described two frequencies for the stellar flare activity. These frequencies are defined as given by Equations (3) and (4):

\begin{center}
\begin{equation}
N_{1}~=~\Sigma n_{f}~/~\Sigma T_{t}
\end{equation}
\end{center}

\begin{center}
\begin{equation}
N_{2}~=~\Sigma P~/~\Sigma T_{t}
\end{equation}
\end{center}
where $\Sigma n_{f}$  is the total flare number detected in the observations, and $\Sigma T_{t}$ is the total observing duration, while $\Sigma P$ is the total equivalent duration obtained from all the flares. In this study, $N_{1}$ frequency was found to be 0.01351 $h^{-1}$, while $N_{2}$ frequency was found to be $0.00006$.

\section{Results and Discussion}
The results obtained from the analyses of short cadence data in the Kepler Mission Database indicated that KIC 9761199 exhibits high level chromospheric activity. However, it needs to identify which component is active, then, it should be compared with its analogue to be certain about activity level of the system. All these need to determine the physical parameters of the components, firstly.

Although all the variations apart from the geometrical effect, such as flare activity, were removed from the short cadence data taken from the database, neither the eclipses nor any other variation could be seen clearly in these eliminated data, as it is seen from the upper panel of Figure 1. To use in the light curve analysis, we computed the averages phase by phase with interval of 0.001 for the eliminated short cadence data. As it can be seen from Figure 2, both the primary and secondary minima and also a sinusoidal variation due to the rotational modulation are clearly visible in these averaged data. Here is an important point that the amplitude difference between the primary and secondary minima can be visible. This is important, because the orbital period of the system is a controversial point in the literature \citep{Wat06, Cou11}. However, this figure demonstrates that the orbital period of the system is 1.3839980 day, not 0.692031 day. According to this result, first of all, KIC 9761199 is a double object. There are two different minima, because of this, the secondary object can not be a planet, it must be a star. If it was a planet, there would not be any  secondary minimum in the light curve due to the contrast effect between a star and a planet.

In the literature, although there are a few approaches using some calibrations for the physical parameters of the components, there is no complete light curve analysis for KIC 9761199. In this purpose, the light curve obtained from the detrended short cadence data was analyzed, using the PHOEBE V.0.32 software \citep{Prs05}, which is employed in the 2003 version of the Wilson-Devinney Code \citep{Wil71, Wil90}. Although the spectral type was given as M1V for KIC 9761199 \citep{Mui12}, there are several temperature values mentioned for this system. According to the de-reddened colors $(H-K)_{\circ} = 0^{m}.15$ and $(J-K)_{\circ} = 0^{m}.58$ of the system, the temperature of the primary component was taken 4060 K as given by \citet{Cou12}. In the light curve analysis with the PHOEBE V.0.32 software, the temperature of the secondary component was found to be 3891$\pm$1 K. The mass ratio of the system ($q$) was found to be 0.69$\pm$0.01, while the inclination ($i$) of the system was found to be $77^\circ.44\pm0^\circ.01$.

In addition, as it is seen from Figure 1 and the upper panel of Figure 2, the light curve exhibits a remarkable sinusoidal variation at out-of-eclipses. This variation could be caused by the effects of tidal distortion or the mutual heating of the components themselves. All these cases were included to the analysis. For this aim, the albedos ($A_{1}$ and $A_{2}$), the gravity-darkening coefficients ($g_{1}$ and $g_{2}$) of the components were computed depending on their temperatures and their dimensionless potentials ($\Omega_{1}$ and $\Omega_{2}$) were also derived in the analysis. However, it was seen from the initial results that the synthetic light curve obtained with these parameters did not fit the observations. In this point, considering both the temperatures and the fractional radii of the components, and also considering the flare activity, we assumed that this sinusoidal variation is caused by the rotational modulation due to the stellar cool spots. Because of this, the sinusoidal variation was modelled with two cool spots on the primary component in the light curve analysis. The spots are separated by about $180^\circ$ longitudinally, while one of them is located about the latitudes of $47^\circ.0\pm0^\circ.2$ and the other one is located about the latitudes of $30^\circ.0\pm0^\circ.2$. Here, it must be noted that it was assumed the cool spots locate on the primary component. However, it is likely they can be located on the secondary component, too. Someone can easily reach an acceptable solution, astrophysically. A strong clue for the answer of this quandary is come from the discussion of flare activity in the later part of the text. The 3D model of Roche geometry for the components and also the cool spots on the primary component is shown in Figure 3 for different phases. 

Considering both the component temperatures and fractional radii found from the light curve analysis, using the calibrations given by \citet{Tok00}, and also Kepler's third law, we tried to estimate the absolute parameters of the components. The mass of the primary component found be 0.57$\pm$0.01 $M_{\odot}$ while it was found to be 0.39$\pm$0.02 $M_{\odot}$ for the secondary component. In addition, the radius of the primary component was computed as 0.62$\pm$0.05 $R_{\odot}$, while that of the secondary component was computed as 0.56$\pm$0.05 $R_{\odot}$.

When it is considered the results of the analysis, it is seen that both the locations of the stellar spots are not changed on the primary component along 289.82931 day (9.66 months), during all the observing season. However, this is a serious moot point for the astrophysical perspective, because a spotted area evolves in 2 or 3 months at most in the solar case  \citep{Ger05}. On the other hand, according to \citet{Hal89} and \citet{Ger05}, it is well known that the spotted areas on the active components of some RS CVn binaries, such as V478 Lyr, can keep their shapes and locations about two years. Therefore, the behaviour of the cool spot activity observed in the case of KIC 9761199 is not inconsistent event in respect to the stellar spot activity phenomenon. It must be noted that although we stated the word spot, actually it refers to an active area, in which several spots can appear and disappear in time. Here, the question to be answered is why the locations of these areas are stable for 289.82931 days. This is a discussion about the differential rotation of the components.

We detected 94 flare events from KIC 9761199, and calculated some parameters for each flare, such as flare frequencies. \citet{Yol16} recently resolved the chromospheric activity nature of a different system, FL Lyr. They found both flare frequencies as $N_{1}=$ 0.41632 $h^{-1}$ and $N_{2}=$ 0.00027 for that system. Comparing the flare frequencies of KIC 9761199 with those of FL Lyr, it is clearly seen that FL Lyr exhibits flare more frequently than KIC 9761199, and also its flares are more powerful than those of KIC 9761199. Comparing KIC 9761199's flare frequency values with those found for UV Ceti type flare stars from spectral type dK5e to dM6e, the flare energies obtained from KIC 9761199 are remarkably lower than those obtained from UV Ceti stars. For instance, the observed flare number per hour for AD Leo was found to be $N_{1}=$ 1.331 $h^{-1}$, while it was found to be $N_{1}=$ 1.056 $h^{-1}$ for EV Lac. Moreover, $N_{2}$ frequency was found to be $0.088$ for EQ Peg, while it was found to be $N_{2}=$ 0.086 for AD Leo \citep{Dal11}. As it is clearly seen from these results, the flare frequencies of KIC 9761199 are remarkably small. However, it is well known from \citet{Dal11} that the flare frequency can dramatically changes from one season to the next for some stars, such as V1005 Ori, EV Lac, etc. In this case, there could be some changes in the flare frequency and the flare behaviour of KIC 9761199 in the next observing seasons. In addition, these results found from the flare frequency analyses reveal why any flare had not been detected from this system by any ground based telescope before the Kepler Mission. If $N_{2}$ frequency is especially considering, it will be understandable, because, $N_{2}$ frequency indicates that the flare events occurring on the active component of the system are so weak to observe by any ground based telescope.
 
The $Plateau$ value was found to be 1.951$\pm$0.069 s from the OPEA model derived from the variation of the flare equivalent duration distributions on the logarithmic scale versus flare total durations for 94 flares. However, \citet{Yol16} found the $Plateau$ value as 1.232$\pm$0.069 s for FL Lyr. Moreover, \citet{Dal11} computed the $Plateau$ values for some UV Ceti type stars. They gave the $Plateau$ values as 3.014 s for EV Lac ($B-V=1^{m}.554$) and 2.935 s for EQ Peg ($B-V=1^{m}.574$), and also it is 2.637 s for V1005 Ori ($B-V=1^{m}.307$). As it is seen that the maximum flare energy detected from KIC 9761199 is remarkably smaller than the maximum energy level obtained from UV Ceti type single flare stars. The $Plateau$ value of this system a bit approaches just to the value obtained from V1005 Ori. It must be noted that \citet{Dal11} found that the $Plateau$ value is always constant for a star, while it is changing from one star to the other depending on their $B-V$ color indexes. The authors defined the $Plateau$ value as the energy saturation level for the flare mechanism occurring on the star. As a result, the flare activity should be occurred on the primary component due to its $B-V$ color index. If it would be the secondary component, the $Plateau$ value of KIC 9761199 was seen a bit incompatible according to the analogue of the secondary component.

Using the regression calculations, the $half-life$ value was found to be 1014.0 s from the OPEA model for KIC 9761199. In the case of FL Lyr, it is 2291.7 s \citep{Yol16}. It means that a flare occurring on the FL Lyr can reach the maximum energy level when the flare total duration reaches $n\times 38$ minutes, while it takes $n\times 17$ minutes for KIC 9761199. Here $n$ is a constant depending on the OPEA function of a star \citep{Spa87, Daw04, Mot07}. In other words, the parameter $half-life$ refers a minimum duration limit for a flare reached maximum energy. In this perspective, the flares, whose total times are shorter than $n\times 17$ minutes, can never reach the $Plateau$ value obtained from the OPEA model derived for the flares of KIC 9761199. This value is a few times higher than those obtained from single dMe stars. Because, in the case of single dMe stars, for example, it was found to be 433.10 s for DO Cep ($B-V=1^{m}.604$), and 334.30 s for EQ Peg, while it is 226.30 s for V1005 Ori \citep{Dal11}. It means that in the case of the stars such as EQ Peg, V1005 Ori and DO Cep, the flares can reach the maximum energy level at their $Plateau$ value, when their total durations reach about $n\times 5$ minutes, while it needs $n\times 17$ minutes for KIC 9761199.

On the other hand, maximum flare rise time ($T_{r}$) obtained from the flares of eclipsing binary KIC 9761199 was found to be 1118.098 s, while maximum flare total time ($T_{t}$) was found to be 6767.72 s. However, these values are $T_{r}=5179.00$ s and $T_{t}=12770.62$ s for FL Lyr. As a result, FL Lyr flare time scales are larger than those obtained from KIC 9761199, which is in agreement with the results found by \citet{Dal11} for the single flare stars from dMe. In the light curve analysis, the primary component is assumed as the chromospherically active component and its color index was found to be $(B-V)=1^{m}.303$. However, the ($B-V$) color index of the active component of FL Lyr is $0^{m}.74$. Consequently, the chromospherically active component of KIC 9761199 is cooler than that of FL Lyr. In this case, the flare time scales of FL Lyr must be larger than the other according to \citet{Dal11}.

Finally considering the $Plateau$ value, flare frequencies and also flare time scales, it is clearly seen that the flare activity level of KIC 9761199 is clearly lower than almost all the UV Ceti type stars, as it is in the case of FL Lyr. This result about the chromospheric activity level of the system is also in agreement with the result found by \citet{Dal11}. The author indicated that all these parameters derived from the OPEA model of a star are depend on just the ($B-V$) color index of that star. In other word, the OPEA model parameters increase or decrease according to variation of the ($B-V$) color index from a star to the other. As a matter of course, the OPEA model parameters derived for KIC 9761199 were found as expected values for the ($B-V$) color index of the primary component, not the secondary component.

In this point, it could be better to recapitulate that comparing the flare activity nature seen from KIC 9761199 indicated that the assumption of "the chromospherically active star is the primary component of the system" is correct. In addition, the statistical analyses of the OPEA model demonstrated that there is only one star exhibiting the flare activity in the system. Because, according to the statistical analyses, the probability value found to be $p-value<0.001$. It means that there is no other function to model the distributions of flare equivalent durations \citep{Mot07, Spa87}. Therefore, there is only one star exhibiting flare activity, in front of us. Considering the classical theory of solar flare \citep{Ger05}, the star exhibiting flare activity and cool spot activity must be the same star.

According to \citet{Dal11}, it is a contentious issue which parameter, the $Plateau$ value or the flare frequencies ($N_{i}$), is the best indicator for the activity level. In general, the chromospheric activity level of a system depends on some parameters such as especially stellar rotation velocity. The rotation velocity depends on generally the stellar age or being a component of a close binary. Because of this, we discussed KIC 9761199 for both cases. Firstly, the age of KIC 9761199 is given as 0.77 Gyr \citep{Wal13}, while the age of FL Lyr is given between 3.05 Gyr and 15.25 Gyr in the literature. Considering \citet{Sku72}'s law, it would be expected that the activity level of KIC 9761199 is clearly higher than that of FL Lyr. Secondly, in this study we computed the radii of the components as $R_{1}=0.62$ $R_{\odot}$ and $R_{2}=0.56$ $R_{\odot}$, and the semi-major axis as 5.16 $R_{\odot}$. However, the radii of the FL Lyr components are given as $R_{1}=1.283$ $R_{\odot}$, $R_{2}=0.963$ $R_{\odot}$, and its semi-major axis is given as $a=9.17$ $R_{\odot}$ by \citet{Eke14}. According to these absolute parameters of the systems, the activity level of KIC 9761199 should be higher than FL Lyr activity level. If the flare frequencies ($N_{i}$) are considered, the flare frequencies of FL Lyr are higher than KIC 9761199. However, if the $Plateau$ values are considered, the $Plateau$ value of KIC 9761199 is clearly higher than that of FL Lyr. According to the semi-major axis of the KIC 9761199, the components are closer to the each other than the status of FL Lyr's components. Maybe this is why the $Plateau$ value is higher in the case of KIC 9761199. However, there is still an unsolved point yet. Why the flare frequencies are higher in the case of FL Lyr? The similar phenomenon is also common among the UV Ceti type single stars \citep{Dal10, Dal11}.

As a result, according to possible masses and radii of the components, KIC 9761199 must be a low-mass and a close binary system. In addition, according to the OPEA model parameters, just one component of the system is chromospherically active star. Both the age given in the literature and the proximity of the components can help to keep the activity level as possible as high. However, the active component exhibits intense flare activity, while it exhibits quiescent spot activity. It does not mean that the cool spot activity level is low in this system. It is possible that the active component could be covered by some large spots spread to the all surface. In this case, there would be no rapid light variation in the light curve out-of-eclipses contrary to the case of FL Lyr. To reveal and understand the nature of the cool spot activity in this system, it needs to track the sinusoidal light variation due to the rotational modulation in the light curve. However, these observations will be very difficult for the ground based telescope due to the sensitivity problems. In addition, there is no detailed spectroscopic observation of the KIC 9761199 in the literature. Therefore, it is also needed for the future studies.

\section*{Acknowledgments} The authors thank the referee for useful comments that have contributed to the improvement of the paper.

\clearpage

\begin{figure*}
\includegraphics[width=12cm]{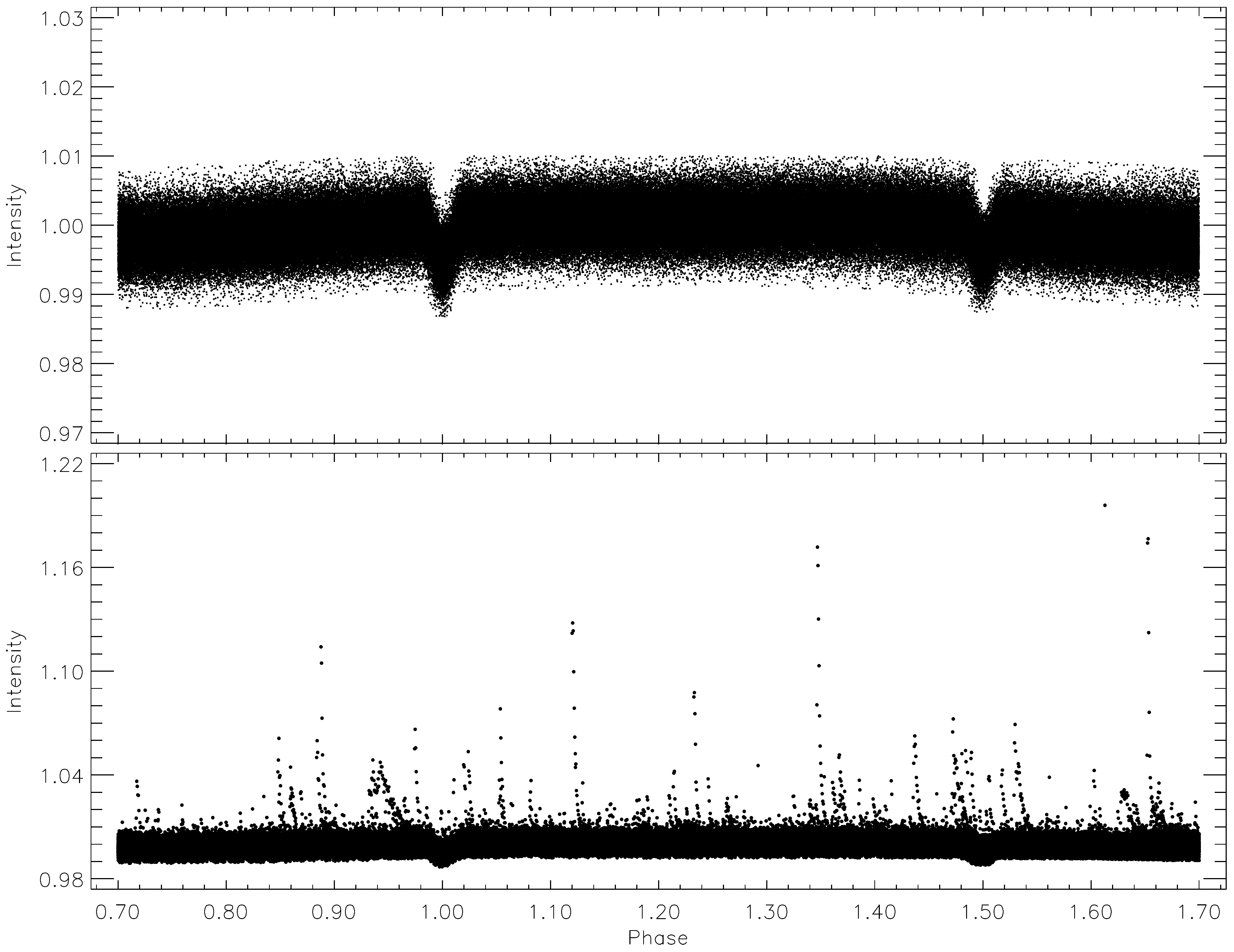}
\vspace{0.1 cm}
\caption{All the light curve of KIC 9761199 obtained from the detrended short cadence data taken from the Kepler Mission database. In the bottom panel, the light curve was plotted with the flare activity, while the light curve was plotted without the flare activity in the upper panel.}
\label{Fig. 1.}
\end{figure*}

\begin{figure*}
\includegraphics[width=12cm]{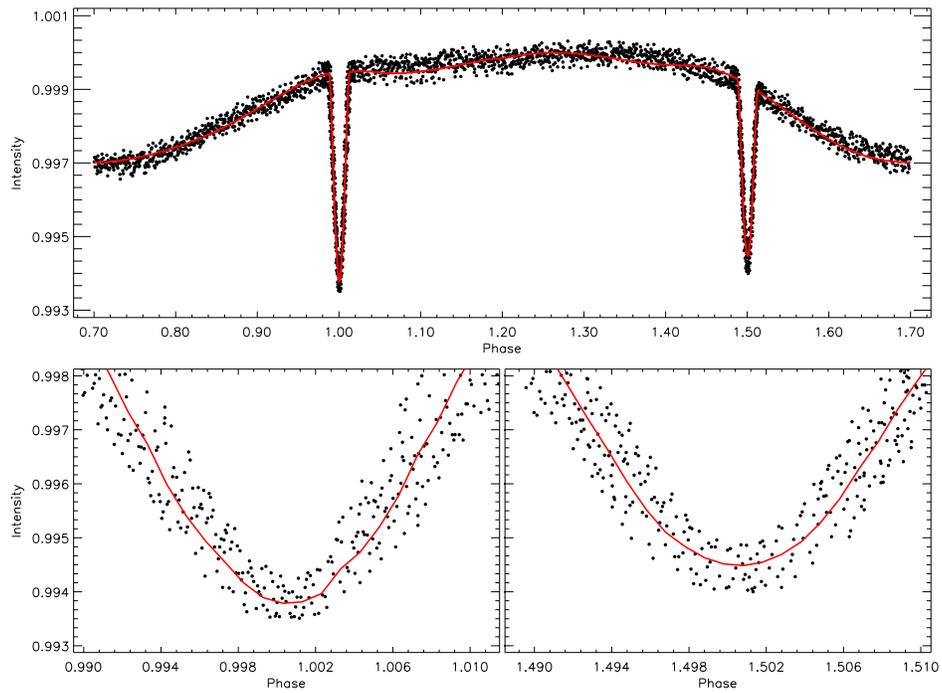}
\vspace{0.1 cm}
\caption{The observed (filled circles) and synthetic (red smooth line) light curves obtained from the averaged short cadence data acquired from HJD 24 54964.50251 to 24 56424.01145. In the bottom panels of the figure, the minima are plotted in the wide plane to better view.}
\label{Fig. 2.}
\end{figure*}

\begin{figure*}
\includegraphics[width=12.4cm]{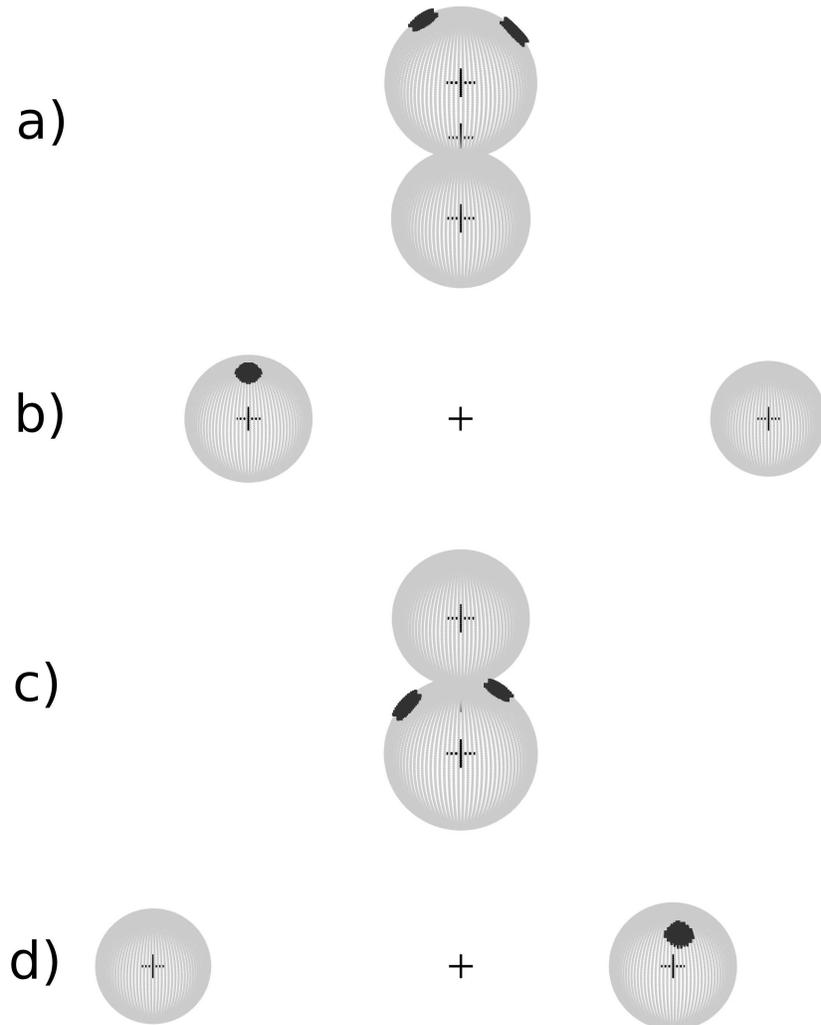}
\vspace{0.1 cm}
\caption{The 3D model of Roche geometry and the cool spots depending on the parameters obtained from the light curve analysis is shown for different phases, such as a) 0.00, b) 0.25, c) 0.50, d) 0.75.}
\label{Fig. 3.}
\end{figure*}

\begin{figure*}
\includegraphics[width=12.4cm]{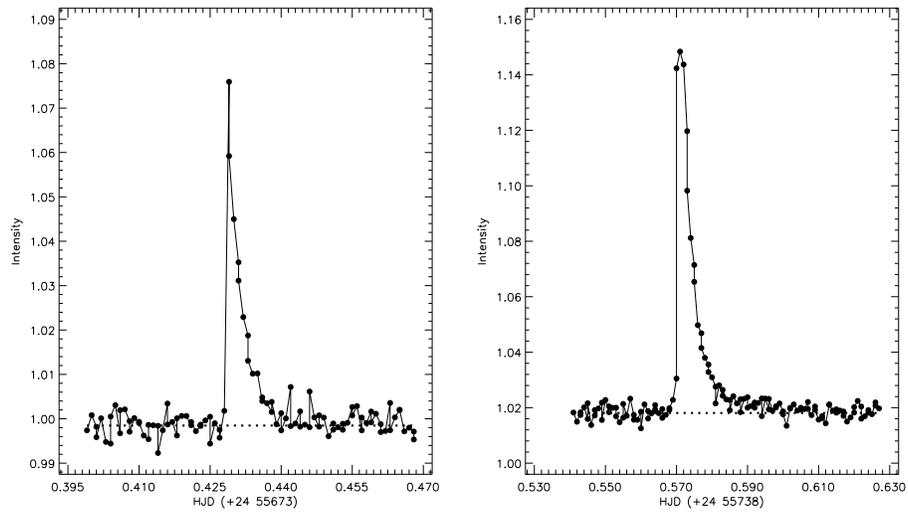}
\vspace{0.1 cm}
\caption{Two different samples for the flare light variations. In the figure, the filled circles represent the observations, while the dotted lines represent the synthetic light curve obtained by the light curve analysis, which was assumed as the quiescent levels for each flare.}
\label{Fig. 4.}
\end{figure*}

\begin{figure*}
\includegraphics[width=12.4cm]{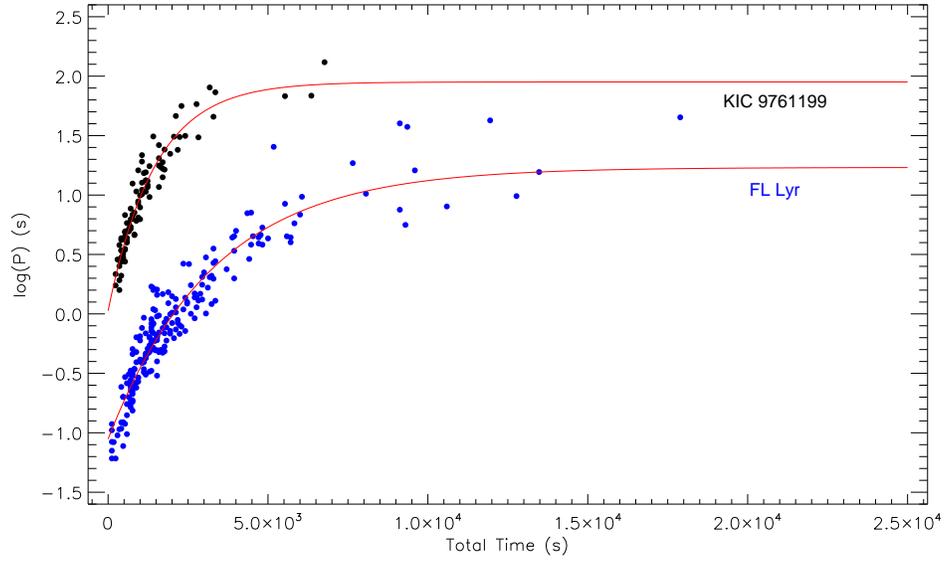}
\vspace{0.1 cm}
\caption{The OPEA model derived from 94 flares detected in the available short cadence data of KIC 9761199 in the Kepler Mission Database is shown, comparing to the OPEA model obtained from the data of well known eclipsing binary, FL Lyr \citep{Yol16}. In the figure, the filled circles represent the calculated $log(P)$ values from observation, while the lines represent the OPEA models.}
\label{Fig. 5.}
\end{figure*}

\clearpage

\setcounter{table}{0}
\begin{table*}
\centering
\footnotesize
\begin{minipage}{15.3 cm}
\caption{The parameters obtained from the light curve analysis of KIC 9761199.}
\begin{tabular}{lr}
\hline
Parameter	&	Value	\\
\hline			
$q$	&	0.69$\pm$0.01	\\
$i (^\circ)$	&	77.4$\pm$0.1	\\
$T_{1} (K)$	&	4060 (fixed)	\\
$T_{2} (K)$	&	3891$\pm$1	\\
$\Omega_{1}$	&	8.96$\pm$0.03	\\
$\Omega_{2}$	&	7.47$\pm$0.03	\\
$L_{1}/L_{T}$	&	0.61$\pm$0.02	\\
$g_{1}, g_{2}$	&	0.32 (fixed)	\\
$A_{1}, A_{2}$	&	0.50 (fixed)	\\
$x_{1},_{bol}, x_{2},_{bol}$	&	0.696, 0.686 (fixed)	\\
$x_{1}, x_{2}$	&	0.709, 0.700 (fixed)	\\
$< r_{1} >$	&	0.121$\pm$0.001	\\
$< r_{2} >$	&	0.109$\pm$0.001	\\
$Co-Lat_{Spot I}$ ($rad$)	&	0.82$\pm$0.03	\\
$Long_{Spot I}$ ($rad$)	&	1.71$\pm$0.03	\\
$R_{Spot I}$ ($rad$)	&	0.18$\pm$0.01	\\
$T_{f Spot I}$	&	0.94$\pm$0.01	\\
$Co-Lat_{Spot II}$ ($rad$)	&	0.52$\pm$0.03	\\
$Long_{Spot II}$ ($rad$)	&	4.71$\pm$0.03	\\
$R_{Spot II}$ ($rad$)	&	0.16$\pm$0.01	\\
$T_{ f Spot II}$	&	0.95$\pm$0.01	\\
\hline
\end{tabular}
\end{minipage}
\end{table*}
\smallskip

\setcounter{table}{1}
\begin{table*}
\centering
\footnotesize
\begin{minipage}{15.3 cm}
\caption{The flare parameters computed from KIC 9761199's the available short cadence data in the Kepler Mission Database.}
\begin{tabular}{ccccc}
\hline
Flare Time	&	$P$	&	$T_{r}$	&	$T_{d}$	&	Amplitude	\\
(+24 00000)	&	(s)	&	(s)	&	(s)	&	(Intensity)	\\
\hline									
55648.62149	&	10.66427	&	294.25248	&	882.75744	&	0.03506	\\
55654.62302	&	3.46467	&	117.70704	&	411.95088	&	0.01650	\\
55656.57790	&	2.75499	&	58.84877	&	470.80915	&	0.01424	\\
55657.21749	&	4.84865	&	117.69840	&	411.95952	&	0.02263	\\
55668.99378	&	6.77842	&	117.69840	&	411.95261	&	0.03122	\\
55671.88592	&	5.40725	&	235.39594	&	470.81088	&	0.02135	\\
55673.42871	&	21.66523	&	58.85741	&	1000.46189	&	0.07746	\\
55682.84894	&	23.99511	&	353.10384	&	1824.38525	&	0.02916	\\
55690.01663	&	7.87550	&	176.55667	&	823.90608	&	0.02173	\\
55690.40829	&	3.41761	&	176.55667	&	235.39680	&	0.01292	\\
55691.66023	&	2.52343	&	176.54717	&	176.55667	&	0.01662	\\
55691.88910	&	2.80804	&	58.75200	&	294.62400	&	0.01503	\\
55693.20098	&	8.51286	&	235.41408	&	706.20768	&	0.02361	\\
55694.47881	&	13.31688	&	353.10384	&	823.92336	&	0.02229	\\
55694.63820	&	2.09756	&	176.55581	&	235.40544	&	0.01227	\\
55696.43438	&	6.47066	&	353.11248	&	588.51706	&	0.01321	\\
55697.18023	&	15.54061	&	353.10384	&	823.90522	&	0.04237	\\
55698.16517	&	4.10949	&	117.69840	&	470.81002	&	0.01660	\\
55701.69145	&	30.56748	&	411.96125	&	2412.88330	&	0.03340	\\
55710.67574	&	11.65687	&	176.53853	&	1059.31757	&	0.02503	\\
55710.97272	&	4.24631	&	176.55581	&	294.25421	&	0.02309	\\
55716.28700	&	18.88827	&	294.25334	&	1412.41018	&	0.04641	\\
55719.31808	&	10.14314	&	176.55494	&	882.76003	&	0.02788	\\
55722.99896	&	11.08245	&	235.39507	&	823.91818	&	0.03130	\\
55724.14396	&	9.62495	&	176.54630	&	1118.16202	&	0.02696	\\
55724.95043	&	4.94587	&	117.68890	&	411.96816	&	0.01795	\\
55728.06529	&	56.09567	&	235.39507	&	2059.76563	&	0.07023	\\
55730.64953	&	13.13169	&	176.54544	&	1059.31066	&	0.02468	\\
55731.55544	&	2.16425	&	117.68890	&	117.70618	&	0.01401	\\
55733.11593	&	3.79166	&	117.69754	&	235.40285	&	0.02496	\\
55733.13705	&	1.92040	&	235.39421	&	117.70618	&	0.01899	\\
55733.40201	&	12.70762	&	117.69754	&	941.60362	&	0.03861	\\
55738.40292	&	12.15108	&	176.55408	&	1059.30720	&	0.02349	\\
55738.57116	&	58.17733	&	294.25939	&	2471.71478	&	0.13029	\\
55740.56212	&	11.68516	&	176.54544	&	1412.40586	&	0.02275	\\
55744.06656	&	2.99633	&	235.40198	&	235.40198	&	0.01388	\\
55744.08495	&	1.73474	&	117.68803	&	117.70531	&	0.01503	\\
55754.92383	&	130.99219	&	647.35459	&	6120.36691	&	0.04921	\\
55772.29873	&	19.12241	&	235.40717	&	823.88189	&	0.05312	\\
55773.26593	&	1.58905	&	58.84704	&	294.25421	&	0.01069	\\
55774.78756	&	2.74595	&	176.55062	&	235.38989	&	0.01329	\\
55775.25413	&	6.79622	&	58.85568	&	706.17744	&	0.02373	\\
55775.61308	&	30.78613	&	176.55062	&	2059.71811	&	0.06187	\\
55783.59719	&	80.32655	&	235.38816	&	2942.44013	&	0.17808	\\
55785.26730	&	12.48313	&	117.69408	&	647.34250	&	0.04444	\\
55785.57517	&	16.66883	&	176.54112	&	1530.07402	&	0.03594	\\
55787.04434	&	2.90663	&	235.38816	&	235.40458	&	0.01604	\\
55788.38615	&	5.19716	&	117.69322	&	588.49546	&	0.01873	\\
55788.73897	&	17.18207	&	176.53162	&	1471.22438	&	0.03291	\\
\hline
\end{tabular}
\end{minipage}
\end{table*}
\smallskip

\setcounter{table}{1}
\begin{table*}
\centering
\footnotesize
\begin{minipage}{15.3 cm}
\caption{Continued From Previous Page.}
\begin{tabular}{ccccc}
\hline
Flare Time	&	$P$	&	$T_{r}$	&	$T_{d}$	&	Amplitude	\\
(+24 00000)	&	(s)	&	(s)	&	(s)	&	(Intensity)	\\
\hline		
55791.99062	&	15.28898	&	117.69408	&	1000.42128	&	0.03837	\\
55795.39620	&	2.86915	&	117.70186	&	176.54890	&	0.01564	\\
55819.27527	&	45.55776	&	706.16794	&	2589.28704	&	0.06316	\\
55796.34364	&	3.53016	&	58.85482	&	470.78237	&	0.01915	\\
55801.73670	&	5.81090	&	176.54803	&	411.93360	&	0.02571	\\
55804.52926	&	2.98768	&	117.70099	&	294.24038	&	0.01691	\\
55820.37526	&	14.11267	&	353.08310	&	1353.48883	&	0.01976	\\
55805.77229	&	68.48297	&	235.38557	&	6120.19670	&	0.02991	\\
55806.37235	&	4.65150	&	235.38557	&	588.47990	&	0.01775	\\
55809.49865	&	46.23526	&	882.71856	&	1235.79562	&	0.05201	\\
55810.42632	&	4.35491	&	235.38470	&	353.08570	&	0.01401	\\
55813.21546	&	17.71834	&	176.53853	&	1412.35747	&	0.03683	\\
55814.02734	&	4.33411	&	58.85395	&	353.07619	&	0.02484	\\
55815.19204	&	6.06366	&	58.85482	&	823.86115	&	0.01632	\\
55815.65178	&	3.20557	&	58.85482	&	411.93014	&	0.01947	\\
55825.57891	&	6.81791	&	176.53680	&	588.47386	&	0.02193	\\
55828.40004	&	24.12872	&	176.55408	&	1588.86576	&	0.05481	\\
55832.00784	&	67.89115	&	1118.09894	&	4413.55046	&	0.03464	\\
55842.95719	&	3.41716	&	58.85395	&	353.07187	&	0.01880	\\
55846.62150	&	7.68861	&	117.69840	&	588.47731	&	0.02595	\\
55848.41960	&	5.73742	&	117.68976	&	529.63200	&	0.02068	\\
55852.56067	&	20.40242	&	353.08829	&	1235.78006	&	0.05653	\\
55862.50127	&	31.43598	&	647.31053	&	1765.39910	&	0.03707	\\
55874.87472	&	17.49964	&	176.52557	&	1118.09462	&	0.04550	\\
55878.24750	&	3.98561	&	176.54285	&	411.92237	&	0.01816	\\
55880.92556	&	26.30278	&	176.53421	&	1412.32723	&	0.06571	\\
55881.21230	&	4.94509	&	176.53421	&	411.93965	&	0.02558	\\
55881.32536	&	9.92092	&	176.53507	&	823.85251	&	0.02456	\\
55884.23091	&	6.25840	&	235.37952	&	765.00806	&	0.01940	\\
55886.83678	&	10.71779	&	58.84445	&	823.85338	&	0.03534	\\
55886.96755	&	22.22971	&	294.24125	&	1647.69811	&	0.02492	\\
55888.44757	&	6.25139	&	117.69840	&	588.45658	&	0.02182	\\
55891.30271	&	73.14961	&	470.77546	&	2883.48422	&	0.07367	\\
55896.11193	&	7.12551	&	117.69840	&	706.15584	&	0.02458	\\
55911.04562	&	4.48567	&	235.38038	&	294.24298	&	0.02266	\\
55913.35318	&	4.19466	&	117.69840	&	294.22570	&	0.01940	\\
55913.69850	&	16.36581	&	353.07965	&	1412.31946	&	0.02380	\\
55918.65009	&	31.09451	&	117.69926	&	1294.63142	&	0.08420	\\
55919.97075	&	6.28820	&	176.53594	&	588.47990	&	0.01913	\\
55920.43049	&	30.97705	&	411.92669	&	1647.71280	&	0.04837	\\
55921.79746	&	16.13419	&	176.54458	&	765.00720	&	0.04360	\\
55921.93640	&	5.16064	&	235.38125	&	411.94397	&	0.02153	\\
55922.14755	&	4.60511	&	176.53680	&	647.32522	&	0.01316	\\
55925.62593	&	4.94213	&	176.54544	&	470.77286	&	0.02649	\\
55929.51435	&	8.85268	&	117.70013	&	823.85597	&	0.02607	\\
\hline
\end{tabular}
\end{minipage}
\end{table*}
\smallskip

\setcounter{table}{2}
\begin{table*}
\centering
\footnotesize
\begin{minipage}{15.3 cm}
\caption{The OPEA model parameters by using the least-squares method.}
\begin{tabular}{lcc}
\hline
Parameter	&	Values	&	95$\%$ Confidence Intervals	\\
\hline	 	 	 	 	
$Y_{0}$	&	0.027$\pm$0.054	&	-0.081 to 0.135	\\
$Plateau$	&	1.951$\pm$0.069	&	1.813 to 2.089	\\
$K$	&	0.0007$\pm$0.0001	&	0.0006 to 0.0008	\\
$Tau$	&	1462.9	&	1232.6 to 1799.1	\\
$Half-life$	&	1014	&	854.4 to 1247.1	\\
$Span$	&	1.924$\pm$0.062	&	1.800 to 2.048	\\
\hline	 	 	 	 	 
Goodness of Fit	&	Method	&	Values	\\
\hline	 	 	 	 	 
$R^{2}$	&	 	&	0.9186	\\
$p-value$	&	(D'Agostino-Pearson)	&	0.0011	\\
$p-value$ 	&	(Shapiro-Wilk)	&	0.0009	\\
$p-value$	&	(Kolmogorov Smirnov)	&	0.0009	\\
\hline
\end{tabular}
\end{minipage}
\end{table*}
\smallskip


\clearpage



\begin{thebibliography}{}
\bibitem[Balona (2015)]{Bal15} Balona, L.A., 2015, MNRAS, 447, 2714
\bibitem[Benz (2008)]{Ben08} Benz, A.O., 2008, Living Rev. Solar Phys., 5, 1
\bibitem[Borucki et al. (2010)]{Bor10} Borucki, Wi.J., Koch, D.G., Basri, G., et al., 2010, Sci, 327, 977
\bibitem[Borucki et al. (2011)]{Bor11} Borucki, Wi.J., Koch, D.G., Basri, G., et al., 2011, ApJ, 736, 19
\bibitem[Caldwell et al. (2010)]{Cal10} Caldwell, D.A., Kolodziejczak, J.J., \& Van Cleve, J. E., 2010, ApJL, 713, L92
\bibitem[Coughlin et al. (2011)]{Cou11} Coughlin, J.L., L\'{o}pez-Morales, M., Harrison, T.E., et al.,  2011, AJ, 141, 78
\bibitem[Coughlin \& L\'{o}pez-Morales (2012)]{Cou12} Coughlin, J.L., L\'{o}pez-Morales, M., 2012, AJ, 143, 39
\bibitem[Cutri et al. (2003)]{Cut03} Cutri, R.M., Skrutskie, M.F., van Dyk, S., et al., 2003. The IRSA 2MASS all sky point source catalog. NASA IPAC Infrared Science Archive http://irsa.ipac.caltech.edu/applications/Gator
\bibitem[D'Agostino \& Stephens (1986)]{Dag86} D'Agostino, R B. \& Stephens, M.A., 1986, "Tests for Normal Distribution" in Goodness-Of-Fit Techniques, Statistics: Textbooks and Monographs, New York: Dekker, edited by D'Agostino, R.B., Stephens, M.A.
\bibitem[Dal \& Evren (2010)]{Dal10} Dal, H.A. \& Evren, S., 2010, AJ, 140, 483
\bibitem[Dal \& Evren (2011)]{Dal11} Dal, H.A. \& Evren, S., 2011, AJ, 141, 33
\bibitem[Dal et al. (2012)]{Dal12} Dal, H.A., Sipahi, E., \"{O}zdarcan, O., 2012, PASA, 29, 150
\bibitem[Dawson \& Trapp (2004)]{Daw04} Dawson, B. \& Trapp, R.G., 2004, "Basic and Clinical Biostatistics" (New York: McGraw-Hill), 61
\bibitem[Eker et al. (2014)]{Eke14} Eker, Z., Bilir, S., Soydugan, F., et al., 2014, PASA, 31, 24
\bibitem[Gershberg \& Shakhovskaya (1983)]{Ger83} Gershberg, R. E. \& Shakhovskaya, N. I., 1983, Astrophys. Space Sci., 95, 235
\bibitem[Gershberg et al. (1972)]{Ger72} Gershberg, R.E., 1972, Astrophys. Space Sci., 19, 75
\bibitem[Gershberg (2005)]{Ger05} Gershberg, R.E., 2005, "Solar-Type Activity in Main-Secauence Stars", Springer Berlin Heidelberg, New York, p.53, 191, 360
\bibitem[Green et al. (1999)]{Gre99} Green, S B., Salkind, N.J. \& Akey, T.M., 1999, "Using SPSS for Windows: Analyzing and Understanding Data" (Upper Saddle River, NJ: Prentice Hall), 50
\bibitem[Haisch et al. (1991)]{Hai91} Haisch, B., Strong, K.T., Rodon\'{o}, M., 1991, ARA\&A, 29, 275
\bibitem[Hall et al. (1989)]{Hal89} Hall D.S., Henry G.W., Sowell J.R., 1989. AJ, 99, 396.
\bibitem[Hudson \& Khan (1997)]{Hud97} Hudson, H.S. \& Khan, J.I., 1997, in ASP Conf. Ser. 111, "Magnetic Reconnection in the Solar Atmosphere", ed. R. D. Bentley \& J. T. Mariska (San Francisco, CA: ASP), 135
\bibitem[Ishida et al. (1991)]{Ish91} Ishida, K., Ichimura, K., Shimizu, Y., et al., 1991, Ap\&SS, 182, 227
\bibitem[Jenkins et al. (2010a)]{Jen10a} Jenkins, J.M., Caldwell, D.A., Chandrasekaran, H., et al., 2010a, ApJL, 713, L87
\bibitem[Jenkins et al. (2010b)]{Jen10b} Jenkins, J.M., Chandrasekaran, H., McCauliff, S.D., et al., 2010b, Proc. SPIE, 7740, 77400
\bibitem[Kharchenko (2001)]{Kha01} Kharchenko, N.V., 2001, KFNT, 17, 409
\bibitem[Koch et al. (2010)]{Koc10} Koch, D.G., Borucki, W.J., Basri, G., et al. 2010, ApJL, 713, L79
\bibitem[Lucy (1967)]{Luc67} Lucy, L.B., 1967, Z. Astrophys, 65, 89
\bibitem[Mann et al.(2013)]{Man13} Mann, A.W., Gaidos, E., Ansdell, M., 2013, ApJ, 779, 188
\bibitem[Marcy \& Chen (1992)]{Mar92} Marcy, G.W. \& Chen, G.H., 1992, ApJ, 390, 550
\bibitem[Matijevi\v{c} et al. (2012)]{Mat12} Matijevi\v{c}, G., Pr\v{s}a, A., Orosz, J.A., et al. 2012, AJ, 143, 123
\bibitem[Mirzoyan (1990)]{Mir90} Mirzoyan, L.V., 1990, in IAU Symp. 137, "Flare stars in star clusters, associations and the solar vicinity", proceedings of the 137th IAU Symposium, Byurakan, Armenian SSR, Oct. 23-27, 1989 (A91-55760 24-90). Dordrecht, Netherlands, Kluwer Academic Publishers, 1990, p.1
\bibitem[Motulsky (2007)]{Mot07} Motulsky, H., 2007, "GraphPad Prism 5: Statistics Guide", San Diego, CA: GraphPad Software Inc. Press, 94
\bibitem[Muirhead et al. (2014)]{Mui14} Muirhead, P.S., Becker, J., Feiden, G.A., et al., 2014, ApJS, 213, 5
\bibitem[Muirhead et al. (2012)]{Mui12} Muirhead, P.S., Hamren, K., Schlawin, E., et al., 2012, ApJL, 750, 37
\bibitem[Pettersen (1991)]{Pet91} Pettersen, B.R., 1991, Mem. Soc. Astron. Ital., 62, 217
\bibitem[Pigatto (1990)]{Pig90} Pigatto, L., 1990, in IAU Symp. 137, "Flare stars in star clusters, associations and the solar vicinity", proceedings of the 137th IAU Symposium, Byurakan, Armenian SSR, Oct. 23-27, 1989 (A91-55760 24-90). Dordrecht, Netherlands, Kluwer Academic Publishers, 1990, p.117
\bibitem[Pr\v{s}a \& Zwitter (2005)]{Prs05} Pr\v{s}a, A. \& Zwitter, T., 2005, ApJ, 628, 426
\bibitem[Rodon\'{o} (1986)]{Rod86} Rodon\'{o}, M., 1986, NASSP, 492, 409
\bibitem[Rucinski (1969)]{Ruc69} Rucinski, S.M., 1969, AcA, 19, 245
\bibitem[Skumanich (1972)]{Sku72} Skumanich, A., 1972, ApJ 171, 565.
\bibitem[Slawson et al. (2011)]{Sla11} Slawson, R., Pr\v{s}a, A., Welsh, W.F., et al. 2011, AJ, 142, 160
\bibitem[Spanier \& Oldham (1987)]{Spa87} Spanier, J. \& Oldham, K.B., 1987, "An Atlas of Function", Washington, DC: Hemisphere Publishing Corporation Press, 233
\bibitem[Stauffer (1991)]{Sta91} Stauffer, J.R., 1991, in Proc. "NATO Advanced Research Workshop on Angular Momentum Evolution of Young Stars", ed. S. Catalano \& J.R. Stauffer (Dordrecht: Kluwer), 117
\bibitem[Tokunaga (2000)]{Tok00} Tokunaga, A.T. 2000, in Allen's Astrophysical Quantities, ed. A.N. Cox (Springer), p.143
\bibitem[Urban et al. (1997)]{Urb97} Urban, S., Corbin, T., Wycoff, G., 1997, AAS, 191, 5707
\bibitem[Van Hamme (1993)]{Van93} Van Hamme, W., 1993, AJ, 106, 2096
\bibitem[Walkowicz \&  Basri (2013)]{Wal13} Walkowicz, L.M. \& Basri, G.S., 2013, MNRAS, 436, 1883
\bibitem[Watson (2006)]{Wat06} Watson, C.L., 2006, SASS, 25, 47
\bibitem[Wilson (1990)]{Wil90} Wilson, R.E., 1990, ApJ, 356, 613
\bibitem[Wilson \& Devinney (1971)]{Wil71} Wilson, R.E., Devinney, E.J., 1971, ApJ, 166, 605
\bibitem[Yolda\c{s} \&  Dal (2016)]{Yol16} Yolda\c{s}, E. and Dal, H.A., 2016, PASA, 33, 16
\end{thebibliography}
\end{document}